# Electrical Spin-Flip Current Switching in Layered Diluted Magnetic Semiconductors for Ultralow-Power Spintronics

Lan-Anh T. Nguyen[1,2], Mallesh Baithi[1,3], Tuan Dung Nguyen[1], Krishna P. Dhakal[3], Jeongyong Kim[3], Ki Kang Kim[1,2,3,6], Dinh Loc Duong[4], Philip Kim[5], Young Hee Lee*[1,2,7]

[1] Center for Integrated Nanostructure Physics (CINAP), Sungkyunkwan University, Suwon 16419, Republic of Korea.

[2] Center for Low-Dimensional Quantum Materials, Hubei University of Technology, Wuhan, 430062, China.

[3] Department of Energy Science, Sungkyunkwan University, Suwon 16419, Republic of Korea.

[4] Department of Physics and Astronomy and Frontier Institute for Research in Sensor Technologies (FIRST), University of Maine, Orono, ME 04473, United States.

[5] John A. Paulson School of Engineering and Applied Sciences, Harvard University, Cambridge, MA, USA.

[6] Department of Physics, Sungkyunkwan University, Suwon 16419, Republic of Korea.

[7] Center for 2D Quantum Heterostructures, Institute for Basic Science, Suwon 16419, Republic of Korea.

*Corresponding author. Email: leeyoung@skku.edu

Tuan Dung Nguyen: Current address; Department of Material Science and Engineering, Texas A&M University, College Station, Texas 77840, USA.

**Abstract**

Efficient magnetic switching is a cornerstone for advancing spintronics, particularly for energy-efficient data storage and memory devices[1–3]. Here, we report the electrical switching of spin-flips in V-doped WSe$_2$ multilayers, a van der Waals (vdW)-layered diluted magnetic semiconductor (DMS), demonstrating ultralow-power switching operation at room temperature. Our study reveals unique linear magnetoresistance and parabolic magnetoresistance states, where electrical modulation induces transitions between interlayered ferromagnetic, ferrimagnetic, and antiferromagnetic configurations. We identify an unconventional linear magnetoresistance hysteresis characterized by electrically driven spin flip/flop switching, distinct from conventional random network disorder or linear band-dispersion mechanisms. Applying an electrical voltage across vertical vdW-layered V-doped WSe$_2$ multilayers generates the spin currents at room temperature, driving spin-flip transitions from ferromagnetic to antiferromagnetic states due to a strong spin-transfer torque effect. Notably, the critical current density reaches an ultralow value of ~$10^{-1}$ A/cm$^2$, accompanied by pico-watt power consumption, a record-low spin current density by a six-order-of-magnitude improvement over conventional spintronic devices. These findings establish the V-doped WSe$_2$ multilayer device as a transformative platform for ultralow-power spintronics, underscoring the potential of vdW-layered DMS systems for next-generation energy-efficient spintronic technologies.

**Introduction**

One of the key challenges in modern electronics, particularly in integrating millions of nanodevices, is to minimize power consumption, which often exceeds thermal limits and leads to device failure. An emerging alternative is to exploit the "spin" of electrons rather than their "charge", a field known as spintronics. Spintronics promise to enable energy-efficient quantum computing, advanced data storage, and beyond. Contemporary approaches in spintronics typically involve magnetoresistance (MR) or current-injection devices, where magnetic switching offers the potential for reduced power consumption. The ultimate objective, however, is to achieve control over magnetic order using electrical bias without relying on an external magnetic field. This approach allows voltage-based spin manipulation with minimal current injections, making it suitable for the current technology of integrated circuits.

Spin-transfer torque (STT) and spin-orbit torque (SOT) are fundamental mechanisms in spintronics, allowing the control of magnetic order in magnetic materials through electric currents. STT operates by injecting a spin-polarized current through one magnetic layer, which exerts torque on another magnetic layer. Magnetic tunnel junctions (MTJs), consisting of two ferromagnetic (FM) layers separated by an insulating barrier, are prototypical devices for STT[4–6]. SOT, on the other hand, leverages the spin-orbit coupling in heavy metals (*e.g*., Pt and Ta) or topological insulators to generate a transverse spin current[7–9]. This spin current exerts a torque on an adjacent FM layer, thereby switching its magnetic moment[10]. Both mechanisms are integral to developing advanced spintronic technologies, such as magnetic random-access memory (MRAM) and neuromorphic computing[11–13].

Despite their advantages, STT and SOT face significant challenges in reducing the critical current density ($J_c$) to offer low-power operation. STT devices with simple structure, using the control and reading currents in the same direction, encounter reliability issues at elevated current densities[14,15]. In contrast, SOT devices enable switching with low current densities but require carefully optimized multilayer structures to enhance spin current generation[8,16–18]. Recent advances have focused on topological insulators and 2D van der Waals (vdW) layered materials to enhance spintronic efficiency[9,19–22]. Interfacial engineering and heterostructure design remain central strategies in overcoming these challenges, paving the way for next-generation spintronic technologies.

FM metals enable magnetic switching through STT and SOT mechanisms, but their efficiency requires substantial enhancement for practical applications. High current densities required for their operation limit their feasibility for low-power spintronics. While traditional FM semiconductors provide magnetic switching functionality, their low Curie temperature ($T_c$) renders them inoperable at room temperature. Diluted magnetic semiconductors (DMSs) offer a promising alternative, combining flexibility to integrate diverse semiconductors with magnetic dopants and efficiently modulate spin-polarized states near the Fermi level. However, their inherently low $T_c$ has hindered progress in both research and practical applications[23]. Recent advances in vdW-layered DMSs, particularly V-doped monolayer $WSe_2$ with doping concentrations of 0.1–0.5% V, have demonstrated FM order at room temperature alongside gating capabilities by using magnetic force microscopy[24]. Notably, the injection current in vertical devices can be modulated by magnetic fluctuations, as evidenced by random telegraph noise (RTN) signals in the MR at a low doping concentration of 0.1% V multilayer $WSe_2$[25]. Despite these advancements, achieving electrical switching of magnetic order via MR or SOT/STT, operable at room temperature and without an external magnetic field, remains a significant challenge for using this material to realize ultralow-power spintronics.

This study reports the first observation of MR flip/flop switching via FM to antiferromagnetic (AFM) transition at room temperature under an applied voltage, achieved by introducing a magnetic dopant into a layered semiconductor, specifically 0.3–0.5% V-doped multilayer $WSe_2$. A robust STT vertical device was constructed using homogeneous V-doped $WSe_2$ multilayers sandwiched between two graphene electrodes, replacing conventional MTJ devices with complex and inhomogeneous heterostructures. We directly measured MR hysteresis under external magnetic fields using direct-current (DC) voltage and temperature. The vdW-layered V-doped $WSe_2$ revealed an unusual spin-flip/flop switching mechanism, converting FM to AFM under external magnetic fields, highlighting the intricate interplay between FM and AFM interlayer couplings. Significantly, we demonstrated voltage-driven current flip/flop switching via FM/AFM transitions at room temperature without an external magnetic field. The switch operates at a voltage below 1 mV, achieving a record low on-current density of $\sim 10^{-1}$ A/cm$^2$, six orders of magnitude lower than conventional MTJ devices. Furthermore, the ultralow power consumption of $\sim 10^{-12}$ W is obtained by three to six orders of magnitude lower than existing MTJ technologies, remarking a transformative step toward energy-efficient spintronic devices.

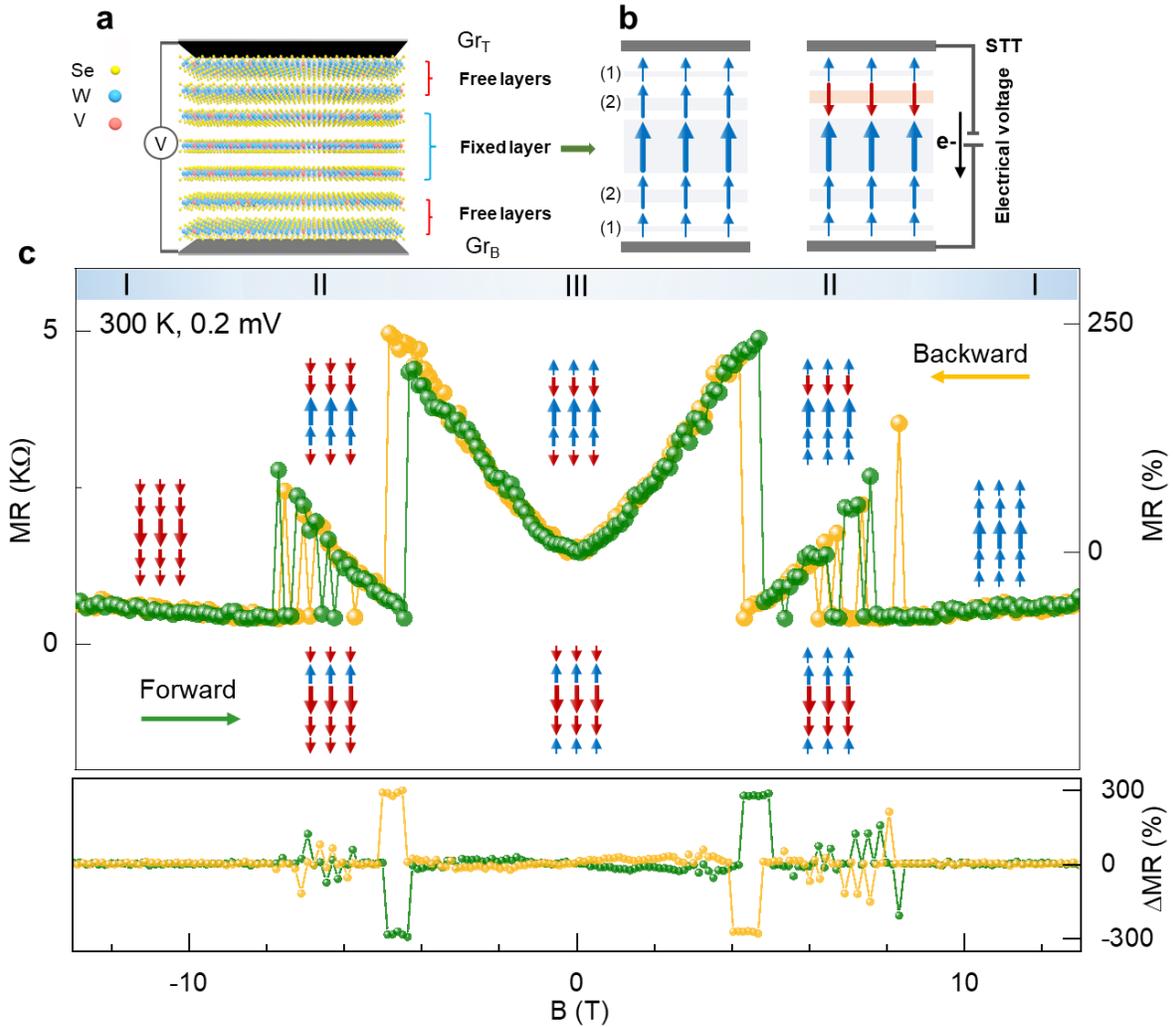

**Figure 1 | Spin flip/flop MR in vertical graphene/V-doped WSe$_2$/graphene device at room temperature. a,** Schematic of the device with multilayered V-doped WSe$_2$ sandwiched between two Gr electrodes. **b,** Simple model of the current-injected STT in the vertical device. **c,** MR hysteresis and different spin states with forward (green) and backward (yellow) directions under external magnetic field (top panel). The relative MR hysteresis is defined as $\Delta MR_{FW}(\%) = MR_{FW}(B)(\%) - MR_{BW}(B)(\%)$ (green) and $\Delta MR_{BW}(\%) = MR_{BW}(B)(\%) - MR_{FW}(B)(\%)$ (yellow) (bottom panel). The data are measured in device D1 at a DC bias of $V = 0.2$ mV and room temperature.

We first investigate the magnetoresistance (MR) of V-doped WSe$_2$ multilayer in the vertical heterostructure (Gr/V-doped WSe$_2$/Gr) (**Fig. 1a**). FM order is established in each layer by introducing a dilute concentration of vanadium as a magnetic dopant[24–26]. Each FM layer is separated by a vdW gap, which acts as an insulating layer analogous to conventional MTJ devices[27–29]. The FM order in each layer is preserved at room temperature at relatively high doping concentrations of 0.3–0.5% V in monolayer WSe$_2$[30], which is confirmed by the spin polarization measurements of circularly polarized photoluminescence at room temperature (**Fig. S1**). Such FM orders still prevail at V-doped WSe$_2$ bilayer and trilayers at 0.3-0.5 %V (**Fig. S2**), implying relatively less-sensitive layer-dependent magnetic order compared to layered magnetic semiconductors like CrI$_3$ and CrSB with FM/AFM alternation[31–34]. Furthermore, under applied vertical bias, interlayer coupling comes into play to establish FM/AFM order between adjacent layers by STT effect (**Fig. 1b**, right panel).

In the vertical Gr/V-doped WSe$_2$/Gr structure, electron transfer occurs from graphene to V-doped WSe$_2$ layers[35,36] compensating holes in V-doped WSe$_2$ to reduce FM order (See **Fig. S3**). Therefore, the magnetic order in the V-doped WSe$_2$ multilayers can be modeled by three main regions: the top and bottom edge layers, which act as two free layers, and the central layer, which serves as a fixed layer (**Fig. 1b**). This spin configuration is further modulated under an applied electrical bias, resembling the STT mechanism in conventional MTJ devices[37]. When spin-unpolarized electrons from the graphene electrodes are injected into the magnetized V-doped WSe$_2$ layers, they become spin-polarized after passing through the first layer (*e.g.*, acquiring an up-spin orientation). These electrons are then reflected at the fixed central layer, switching to a down-spin orientation, ultimately leading to an antiparallel spin configuration. The outer (free) layers gradually lose magnetic order due to the reduction of holes caused by the accumulated electrons, whereas the central (fixed) layer maintains maximum magnetic order, stabilized by the intrinsic carrier density from neutralized charges (**Supplementary Note 1**). Under an applied electrical current, the STT effect induces an AFM order. The similar spin configuration model is valid across the range of sample thicknesses of thin (8 nm, device D1) and thick (40 nm, device D2) samples, although the spin configurations slightly vary in the thicker sample with different applied biases (**Fig. S4**).

**Figure 1c** shows the MR as a function of an applied magnetic field swept up to ±13 T at 300 K with a small electrical bias ($V = 0.2$ mV). The MR curve prominently features spin flip/flop transitions, accompanied by a pronounced MR hysteresis that reflects different spin configurations at room temperature. The MR is defined as MR(%) = 100*[$R$(B) - $R$(0)]/$R$(0) and $\Delta$MR$_{FW(BW)}$(%) = MR$_{FW(BW)}$(B)(%) - MR$_{BW(FW)}$(B)(%), where FW and BW indicate forward and backward sweeps, respectively. Generally, lower resistance corresponds to parallel spin configurations, while higher resistance is associated with antiparallel configurations[27–29]. The magnetic states can be classified into three regimes: ferromagnetic (FM, I), ferrimagnetic (FIM, II), and weak antiferromagnetic (AFM, III) corresponding to their magnetic configurations (**Supplementary Note 2**). For example, at a low temperature (200 K), more distinct magnetic transitions were observed, demonstrating clear spin flip/flop transitions (**Fig. S5**). In general, there are several possible spin configurations of 5-layers with corresponding up-spins/down-spins: 5↑0↓, 4↑1↓, 3↑2↓, 2↑3↓, 1↑4↓, and 0↑5↓ (**Fig. S6**). Such spin configurations can be categorized, depending on five different MR magnitudes, to the corresponding FM, FIM mediated by FM/AFM coupling, and AFM states (**Fig. S7**). FM is favored at the lowest MR. Mixed FM/AFM coupling prevails at intermediate MR, whereas AFM is dominant at high MR. Various MR changes ($R_1$-$R_5$) can be explained in terms of different spin configurations at 200 K (**Fig. S8**).

At high magnetic fields (B > +8 T, backward), the system predominantly exhibits FM order (I), as indicated by the magnetic configuration in **Fig. 1c**. In the intermediate magnetic field range (4∼8 T), FM order weakens and competes with AFM order, mediated by a FIM state (II). This competition leads to fluctuations with RTN signals reaching up to ∼240% ΔMR within a power range of 0.03∼0.08 nW (see **Fig. S9**). As the magnetic field decreases below 4 T, FM order is further suppressed, and AFM order (III) becomes dominant, leading to a sharp increase in MR, with ΔMR∼300%. As the magnetic field moves towards the negative direction, MR change is the opposite of the positive direction with similar MR changes. A similar trend of the MR changes is observed in the forward direction. The MR hysteresis between forward and backward provides concrete evidence of robust FM order in the V-doped $WSe_2$ multilayer, even at high-temperature ranges (**Fig. S10**). These unique MR characteristics, observed in vdW-layered FM semiconductors at room temperature, represent a novel phenomenon not demonstrated in previous studies.

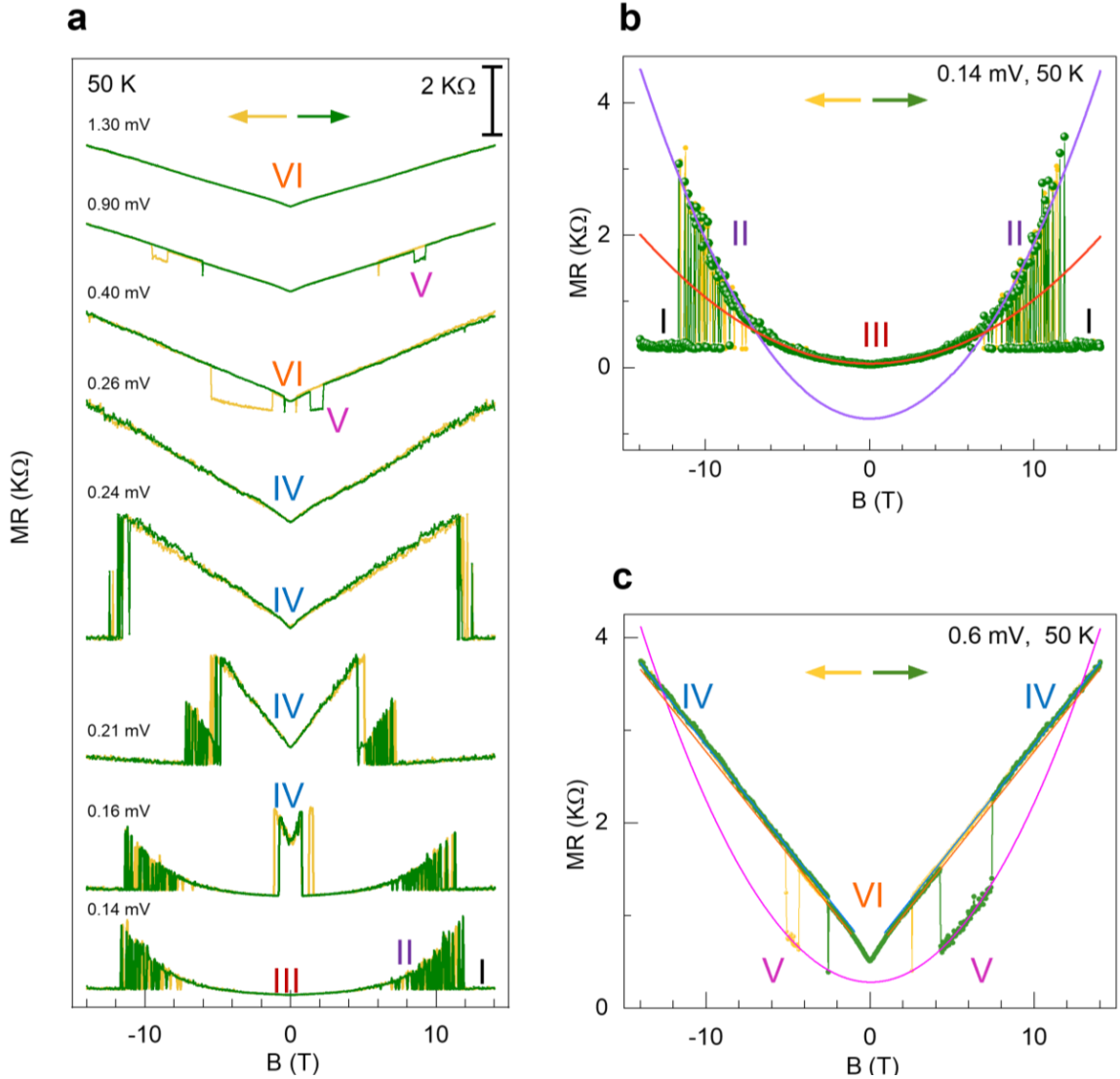

**Figure 2 | Electrical voltage-modulated MR hysteresis in V-doped WSe$_2$. a**, Forward (green) and backward (yellow) MR at DC voltage ranging from 0.14 mV to 1.30 mV at 50 K. Different spin orders are defined by MR: FM states (I, II, III, and V) and AFM states (IV and VI). **b**, Parabolic MR behavior at 0.14 mV is well-fitted with two parabolic curves for corresponding II (purple) and III (red) states. **c**, MR curve at 0.6 mV with linear and parabolic fitting with corresponding two AFM states of IV (blue) and VI (orange) and FM states of V (pink). The data are measured in device D1 at 50 K.

The applied voltage significantly modulates the magnetic order between the layers of the V-doped WSe$_2$ device. **Figure 2a** shows the electrically modulated MR across a range of applied voltages (0.14 − 1.30 mV) at 50 K. At the minimum applied voltage (0.14 mV), the MR follows a classical $R \sim B^2$ relationship within a small magnetic field range (0−6 T). Beyond 6 T, RTN signals dominate, arising from competition between FM states (I) and FIM states (II). The FM states are still preserved in this intermediate field range, as implied by the relatively small MR compared to the significantly larger MR of AFM states. At higher magnetic fields (B > 12 T), the MR stabilizes into a purely FM configuration with low resistance and no fluctuations, similar to the behavior observed in **Fig. 1**.

To distinguish the possible magnetic configurations, the MR curves were fitted using $R \sim a*B^2 + b*B + c$ (**Fig. 2b**). This allows us to identify the two possible magnetic configurations with different coefficients of $a$: (I) the FM state at high magnetic fields (B > 12 T; (II) a FIM state resulting from competing interlayer couplings between FM and AFM orders, with varying magnitudes of spin alignment; and (III) a stable FM state without spin fluctuations. Interestingly, an additional configuration (IV) emerges at low magnetic fields when the applied voltage increases (0.16 − 0.24 mV). Configuration IV is attributed to an AFM state, which exhibits much higher resistance than other configurations and demonstrates greater stability as the applied voltage increases. As the voltage is further elevated (0.21 − 0.26 mV), configurations I, II, and III sequentially disappear. At $V = 0.26$ mV, configuration IV dominates, with a stable linear MR (LMR) spanning the entire magnetic field range (0 − 14 T).

Magnetic switching at higher voltage ranges (0.3 − 1.3 mV) involves a complex dynamic spin-flip process (**Fig. 2a and Fig. S11**). The asymmetric spin-flip switching observed in forward and backward sweeps arises from the presence of distinct inherent magnetic orders. This spin-flip/flop behavior comprises non-saturated LMR states (IV and VI), corresponding to AFM configurations, and parabolic MR states (V), associated with FM configurations. These distinct states are demonstrated through the deconvolution of MR curves (**Fig. 2c**). A second AFM configuration (VI) is also observed, where the MR is smaller than that of AFM configuration IV. At higher voltages ($V \geq 1.3$ mV), the spin-flip effect entirely disappears.

Interestingly, the AFM LMR configurations at $V = 0.26$ mV and $V = 1.30$ mV differ in terms of MR magnitude and slope ($R/B$) (**Fig. S12**). High LMR at $V = 0.26$ mV is attributed to the dominance of magnetic field-induced AFM order, while the lower LMR at $V = 1.30$ mV results from carrier injection dominating the system. These complex spin dynamics and voltage-modulated effects in V-doped WSe$_2$ multilayers are unconventional in vdW-layered magnetic materials. The observed spin flip/flop behavior can be attributed to the interplay between AFM orders induced by STT and the intrinsic FM layers. While this study provides a foundation for understanding these phenomena, further theoretical investigations are required to fully elucidate the detailed mechanisms underlying such complex spin configurations.

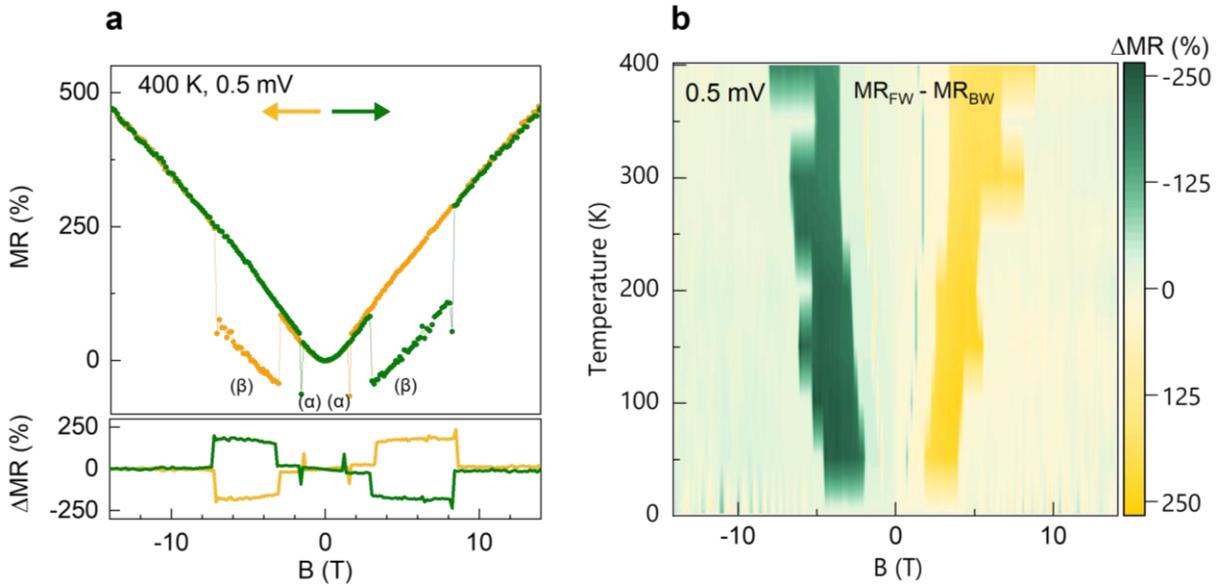

**Figure 3 | Temperature-dependent spin-flip/flop and giant LMR in V-doped $WSe_2$. a,** MR (top panel) with respect to B = 0 and ΔMR hysteresis (bottom panel) at 400 K, $ΔMR_{FW(BW)}$ (%) = $MR_{FW(BW)}$(B) (%) – $MR_{BW(FW)}$(B)(%). **b,** Color map of $ΔMR_{FW}$ hysteresis *vs* B (sweep from +14 T to -14 T) and T (from 2 to 400 K) at a voltage of 0.5 mV. The data are measured in device D1.

**Figure 3a** illustrates how LMR switches through spin-flip mechanisms at a high temperature (400 K) under $V$ = 0.5 mV. In the intermediate voltage range (0.4 − 1.0 mV), STT dominates, leading to AFM LMR states that compete with parabolic FM configuration. This results in two distinct switching behaviors: (i) a sharp spin-flip (α) from the AFM state and (ii) a blunt spin-flip/flop (β) that manifests in the FM state before reverting to the AFM state at higher magnetic fields. These behaviors contrast with the FM state observed at high magnetic fields and low voltage ranges (0.14 − 0.24 mV), as shown in **Fig. 1c** and **Fig. 2a**.

At 400 K, the LMR value reaches as high as ∼500%, with a corresponding ΔMR hysteresis of ∼240%. This high ΔMR hysteresis remains stable across a wide range of temperatures and magnetic fields, as depicted in the color map in **Fig. 3b**. This remarkable stability starkly contrasts conventional FM systems, where magnetic moments typically diminish with increasing temperature[38–42]. For comparison, ΔMR values of 192% at 10 K and 85% at room temperature have been reported for $Fe_3GeTe_2$-based MTJs[43,44]. The negligible variation in ΔMR highlights the robustness of the V-doped $WSe_2$ system at elevated temperatures, making it a promising candidate for applications in MRAM and neuromorphic computing that require stable operation over a wide temperature range (**Fig. 3b**). Furthermore, the coercive fields strengthen with increasing temperature. This phenomenon is attributed to thermally excited carriers associated with V-doping and defect states, such as selenium vacancies[45,46]. These carriers enhance the strong interlayer coupling, reinforcing AFM order and inducing stronger STT effects at higher temperatures.

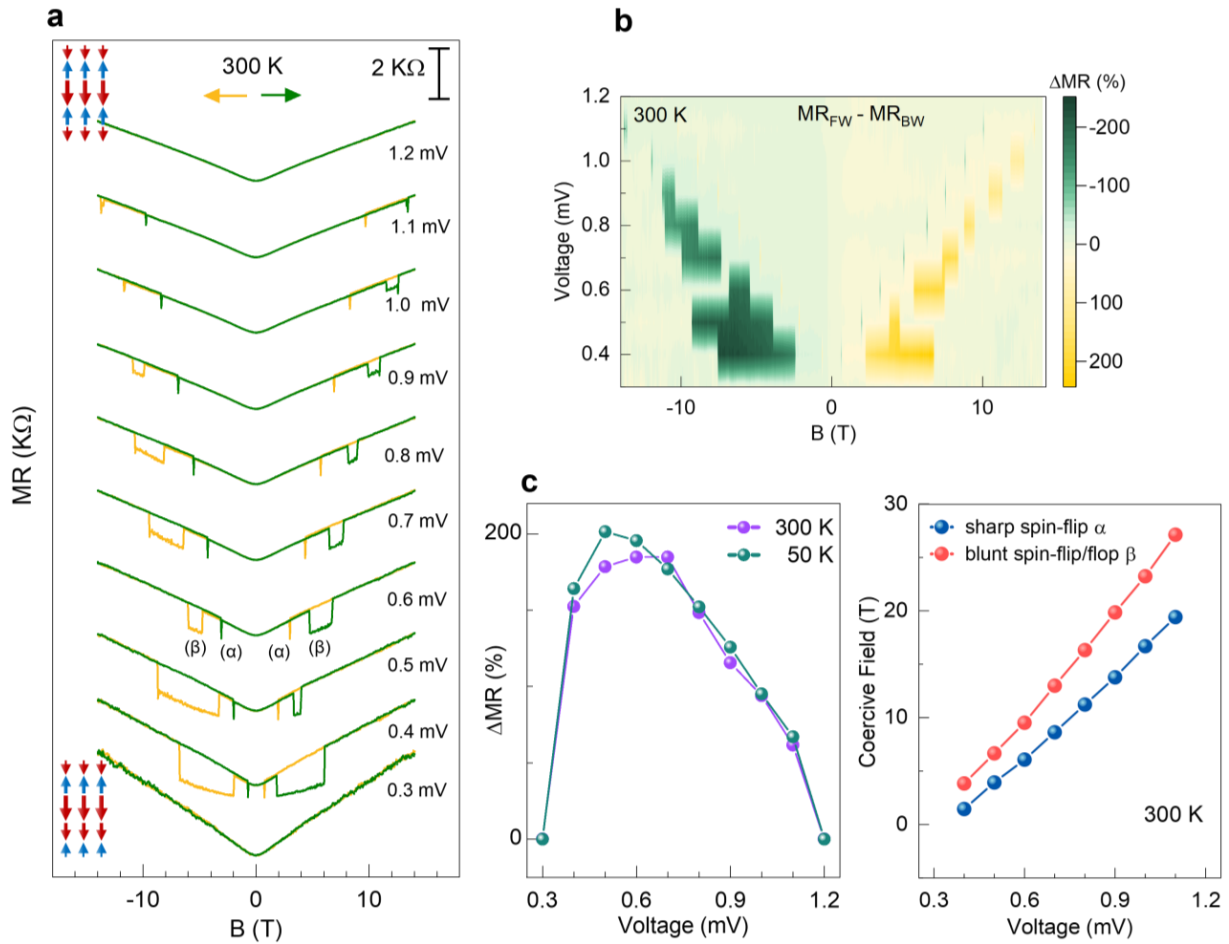

**Figure 4 | Room-temperature voltage-tunable MR in V-doped WSe$_2$ multilayers. a**, Spin flip/flop LMR switching under various DC voltages ranging from 0.3 mV to 1.2 mV at 300 K. **b**, Color map of ΔMR hysteresis *vs B* (sweep from +14 T to -14 T) and voltage V (from 0.4 to 1.2 mV). **c**, ΔMR hysteresis and coercivity field as a function of voltages at 300 K (purple) and 50 K (cyan) for sharp spin-flip switching (α) (blue) and blunt spin-flip/flop (β) switching (red) (right panel). The data are measured in device D1.

We now demonstrate ΔMR switching driven by electrical bias at room temperature. For clarity, this analysis focuses exclusively on the AFM LMR hysteresis in the spin-flip/flop region, excluding RTN signals by selecting a voltage range of 0.3 − 1.2 mV (**Fig. 4a**). As the voltage or injection current increases, spins in the free layers begin flipping, accompanied by an enhancement in the coercive field, as shown in the color map (**Fig. 4b**). AFM spin configuration appears At $V = 0.3$ mV. At $V = 0.4$ mV, FIM configuration emerges, leading to two distinct types of switching: sharp spin-flip (α) and blunt spin-flip/flop (β). At $V = 1.2$ mV, the AFM configuration is fully recovered due to the strong STT.

**Figure 4c** (left panel) highlights the clear electrical switching effect on ΔMR hysteresis as voltage increases. The ΔMR hysteresis reaches a maximum at $V = 0.6$ mV, driven by the dominance of the emerging FIM configuration, and subsequently decreases at higher voltages as the AFM configuration takes over. A similar trend is observed at lower temperatures (**Fig. S13**). The coercive field demonstrates voltage-dependent switching at room temperature, as shown in **Fig. 4c** (right panel), with sharp flip and blunt spin-flip/flop transitions linearly proportional to voltage. These results highlight the critical role of electrical bias in controlling spin dynamics in V-doped WSe$_2$ devices. By tuning the voltage, it is possible to manipulate the magnetic order, transforming from FM/AFM coupling at low voltages to a purely AFM configuration at higher voltages. This tunability offers a promising pathway for engineering magnetic order in spintronic applications.

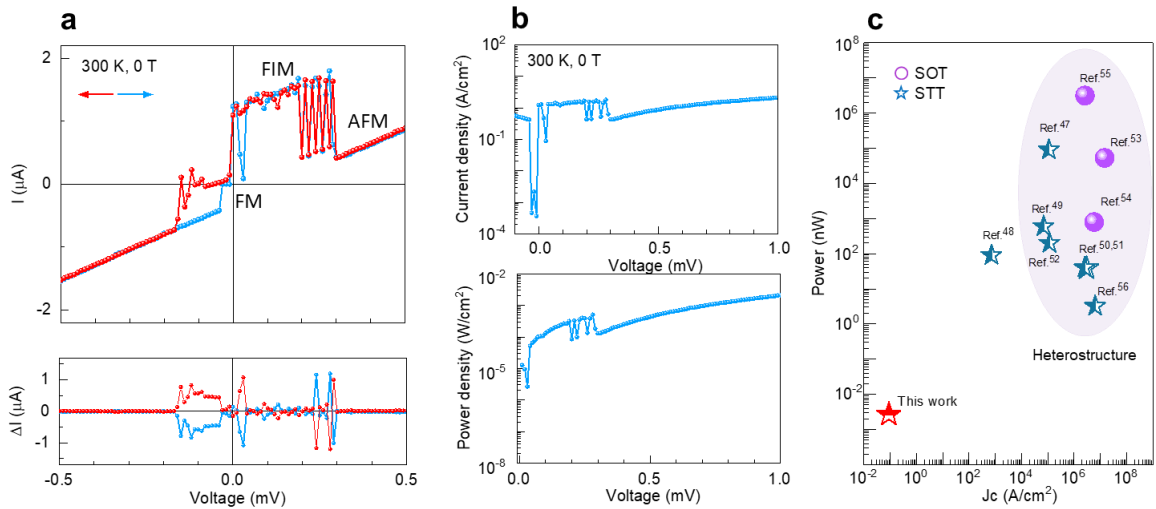

**Figure 5 | *I-V* curves of spin-flip in V-doped WSe$_2$ vertical device at room temperature without external magnetic field. a**, *I-V* curve at $T = 300$ K and $B = 0$ T and the corresponding Δ*I* (current difference between forward (blue) and backward (red) current directions). FM/AFM transition states intermediated by RTN signals are manifested by Δ*I* hysteresis. **b**, Current density and power density as a function of bias. **c**, Roadmap of power vs critical current density.

The development of energy-efficient data storage and memory devices faces significant challenges, primarily due to the high critical current densities required by SOT and STT mechanisms, which typically exceed $10^5-10^6$ A/cm$^2$ [47–57]. **Figure 5a** shows the $I-V$ curve of the V-doped WSe$_2$ device at room temperature and without an external magnetic field. The forward and backward voltage sweeps reveal a clear current hysteresis due to spin polarization modulation, resembling the MR hysteresis observed in **Fig. 1**. The non-negligible spin current hysteresis near zero voltage reflects the intrinsic FM nature of the V-doped WSe$_2$. As the voltage increases, STT induced by strong interlayer coupling drives the system into an AFM state, triggering a spin flip from FM to AFM. At intermediate voltage ranges ($< 0.3$ mV), RTN signals are observed, indicating transitions between FM and FIM states, although the FM state is overall favored. At higher voltages ($> 0.4$ mV), the RTN signals in the spin current disappear, and the spin current becomes linearly proportional to the voltage, corresponding to the stabilized AFM state.

The device exhibits remarkably low current densities and power consumption. The off-current density is as low as $\sim 10^{-1}$ A/cm$^2$ under the FM state, while the on-current density reaches $\sim 1$ A/cm$^2$ during the FM/FIM transition. Even at higher voltages, the current remains low, maintaining the AFM state. The power consumption for the off- and on-spin currents is approximately $\sim 10^{-12}$ W and $\sim 10^{-9}$ W, respectively. As shown in **Fig. 5c**, the critical current density at power-off is $\sim 10^{-1}$ A/cm$^2$, which is six orders of magnitude lower than conventional STT-based MTJ devices and four orders of magnitude lower than CrI$_3$-based SOT devices at low temperatures (30 K) (**Supplementary table 1**). This unusually low spin current density is attributed to robust spin-flip dynamics enabled by the inherent FM order. This remarkable low critical current density and ultralow power underscore the unique properties of our vdW-layered DMSs that the Fermi level of the spin-polarized electrons is located near the valence band edge, offering highly efficient energy transfer through spin-transfer processes.

**Discussion**

*Electrical switching between LMR and Parabolic MR states*

LMR is a well-known phenomenon observed in various material systems, which can generally be categorized into two types. Type I systems exhibit linear band dispersion, such as graphene, Dirac metals, and topological insulators[58–61]. In these materials, LMR arises from the collapse of free carriers at the lowest Landau levels[62]. Type II systems include disordered or inhomogeneous semiconductors with a small concentration of defects (*e.g.*, dopants, vacancies)[63,64], where LMR is explained by random 2D network models[65]. However, electrical magnetic switching behavior in LMR systems has not been reported previously. Here, we identify a new **Type III system** characterized by electrical modulation of magnetic order within the LMR system. In our V-doped WSe$_2$ sample (doping concentration of 0.3% V), the material exhibits characteristics of an inhomogeneous semiconductor, similar to Type II systems. However, we observe clear magnetic switching between LMR and parabolic spin flip/flop states induced by electrical bias, which conventional random-network disorder models cannot explain. This long-range magnetic order, prominent in multilayered magnetic semiconductors, suggests an unusual mechanism that demands further theoretical modeling to explain the behavior of V-doped WSe$_2$ multilayers.

*Spin-current injection favors AFM order to persist in LMR*

The persistence of unsaturated LMR with an AFM configuration, even at high magnetic fields under intermediate voltage ranges (0.4 − 1.1 mV), is unconventional and distinct from typical behaviors in FM systems. This behavior, shown in **Fig. 2a**, contrasts sharply with the FM order observed at high magnetic fields in the absence of electrical bias or under low bias conditions (0.14 − 0.26 mV) (**Fig. 1c** and **Fig. 2a**). The persistence of LMR at high magnetic fields is attributed to strong STT induced by the high electrical bias. In the V-doped WSe$_2$ multilayers, the middle layers act as fixed magnetic layers. Under high electrical bias, injected up-spins are reflected by the fixed layer, converting into down-spins and forming an AFM configuration. This behavior is analogous to current-induced magnetization switching observed in synthetic antiferromagnet-free layers[47]. The fixed nature of the middle layer is corroborated by the consistent FM order at high magnetic fields under low bias, and the transition to AFM order under high bias (**Fig. 2a**). At lower electrical biases (< 0.2 mV), magnetic switching becomes more complex, characterized by the dual parabolic MR curves at low magnetic fields (**Fig. 2b**).

**Conclusion**

We have demonstrated the unique ability to achieve electrically driven magnetic switching in vdW-layered V-doped WSe$_2$ multilayers, paving the way for energy-efficient spintronic devices. Our findings reveal the presence of long-range magnetic order and spin dynamics modulated by electrical bias, resulting in transitions between LMR and parabolic MR states. We further identified an unconventional Type III LMR system, where the switching behavior is driven by electrical modulation of magnetic order rather than conventional mechanisms like random network disorder or linear-band dispersion. Key breakthroughs include the persistence of AFM configurations under high magnetic fields and spin-current flip/flop in intermediate voltage ranges (0.4 − 1.1 mV) facilitated by strong STT. Additionally, we achieved ultralow critical spin current densities ($\sim 10^{-1}$ A/cm$^2$) and power consumption density ($\sim 10^{-6}$ W/cm$^2$), remarking a six-order-of-magnitude improvement over conventional STT-MTJ devices. These results establish V-doped WSe$_2$ as a promising platform for ultralow-power spintronic technologies, including MRAM and neuromorphic computing. Our work highlights the critical role of electrical bias in tuning spin dynamics, enabling transitions from FM to AFM states. This control, combined with the stability of LMR across a wide temperature range, offers practical solutions for robust device operation in energy-efficient data storage and memory systems. Future theoretical and experimental investigations are required to further elucidate the underlying mechanisms of electrical modulation in vdW-layered magnetic semiconductors, particularly the interplay between STT and magnetic order at the atomic scale. This study opens new pathways for designing high-performance spintronic devices with unprecedented energy efficiency and scalability.

**Acknowledgments**

**Funding:** This work was supported by the Institute for Basic Science (IBS-R036-D1). K.K.K. acknowledges support from the Basic Science Research through the National Research Foundation of Korea (NRF), which was funded by the Ministry of Science, ICT and Future Planning, and the Korean government (MSIT) (2022R1A2C2091475 and RS-2024-00439520).

**Author Contributions:** L.-A.T.N. fabricated and characterized devices. M.B., K.K.K., and T.D.N. synthesized V-doped WSe$_2$ single crystals. L.-A.T.N., K. P. D., and J. K. measured circular photoluminescence. D.L.D. and Y.H.L. guided the work. L.-A.T.N., D.L.D., P.K., and Y.H.L. analyzed the data. All authors discussed and wrote the manuscript.

**Competing Interests:** The authors declare that they have no competing interests.

**Data Availability:** All data are available in the main text or the supplementary materials.